\documentclass[conference,letterpaper,10pt]{IEEEtran}

\usepackage{cite}
\usepackage{amsmath,amssymb,amsfonts}
\usepackage{algorithm}
\usepackage{algpseudocode}
\usepackage{graphicx}
\usepackage{textcomp}
\usepackage{xcolor}
\usepackage{array}
\usepackage{booktabs}
\usepackage{multirow}
\usepackage{colortbl}
\usepackage{balance}

\def\BibTeX{{\rm B\kern-.05em{\sc i\kern-.025em b}\kern-.08em
    T\kern-.1667em\lower.7ex\hbox{E}\kern-.125emX}}

\begin{document}

\title{CAFE: \underline{C}onducting-Cell-Aware \underline{A}daptive Differential Encoding for Robust DNN Inference in \underline{Fe}FET-Based Compute-in-Memory}

\author{
    \IEEEauthorblockN{Wenshuai Yao$^{1\dagger}$, Wenyong Zhou$^{2\dagger}$, Hanyong Shao$^{1}$, Zhiyuan Ning$^{1}$, Ru Huang$^{1,3}$, Kechao Tang$^{1,3*}$}
    \IEEEauthorblockA{
        $^1$School of Integrated Circuits, Peking University, Beijing, China\\
        $^2$Department of Electrical and Computer Engineering, The University of Hong Kong, Hong Kong\\
        $^3$Beijing Advanced Innovation Center for Integrated Circuits, Beijing, China\\
         \textsuperscript{\dag}: Equal Contribution. *: Corresponding Author. Email: tkch@pku.edu.cn
    }
}

\maketitle

\begin{abstract}
Ferroelectric field-effect transistor (FeFET)-based compute-in-memory (CIM) promises energy-efficient deep neural network (DNN) inference, but device variation can severely degrade bit-sliced CIM accuracy. Prior works either rely on hardware modifications or treat such variation as generic noise, overlooking a key FeFET asymmetry: negligible off-state leakage makes errors dominated by conducting on-state cells. This asymmetry is amplified by two's-complement sign extension, which maps small negative values to high-significance bit slices and magnifies their errors during shift-and-add reconstruction.
In this paper, we propose CAFE, a conducting-cell-aware adaptive differential encoding framework built upon this FeFET-specific asymmetry. CAFE introduces C\(^2\)AM, a bit-significance-weighted conducting-cell activity metric that predicts output deviation and inference accuracy degradation under FeFET device variation. Guided by C\(^2\)AM, CAFE applies differential weight and activation encoding to eliminate sign-extension-induced conducting-cell amplification. To balance robustness and cost, CAFE performs type-wise encoding enumeration, layer-wise NSGA-II search, and post-GA pruning to allocate robust encodings where the robustness-cost trade-off is favorable.
Across representative vision and language workloads, CAFE reduces C\(^2\)AM by up to 28.6$\times$; in the main deployment study, it improves accuracy over two's-complement CIM by more than 87 percentage points and approaches full-differential robustness at 1.86--2.78$\times$ normalized cost, compared with 3.06$\times$ for uniform full-differential encoding.
\end{abstract}


\section{Introduction}

Deep neural networks (DNNs), especially modern Transformer-based models~\cite{vaswani2017attention}, have become dominant workloads across vision, language, and recommendation tasks. Their inference cost is largely driven by dense matrix-vector and matrix-matrix multiplications, which require frequent data movement between memory and compute units. Compute-in-memory (CIM) addresses this bottleneck by performing multiply-accumulate operations directly inside memory arrays, thereby improving energy efficiency and throughput for DNN inference. Among emerging nonvolatile devices, ferroelectric field-effect transistors (FeFETs) are particularly attractive for CIM because of their CMOS compatibility, nonvolatility, and high on/off ratio~\cite{long2019fefet_pim}.

However, FeFET-based CIM inference remains vulnerable to device variation. In bit-sliced CIM, quantized weights and activations are decomposed into bit planes, computed separately in memory arrays, and reconstructed through shift-and-add operations. Small device-level current variations can therefore be magnified by bit significance and propagate into large output deviations. Prior works typically address this problem in two ways: modifying hardware circuits to compensate for variation, or abstracting variation as generic conductance or output noise for noise-aware training. While effective in some settings, these approaches overlook an important device- and representation-dependent error mechanism in FeFET-based CIM.

\begin{figure*}[!t]
\centering
\includegraphics[width=0.95\textwidth]{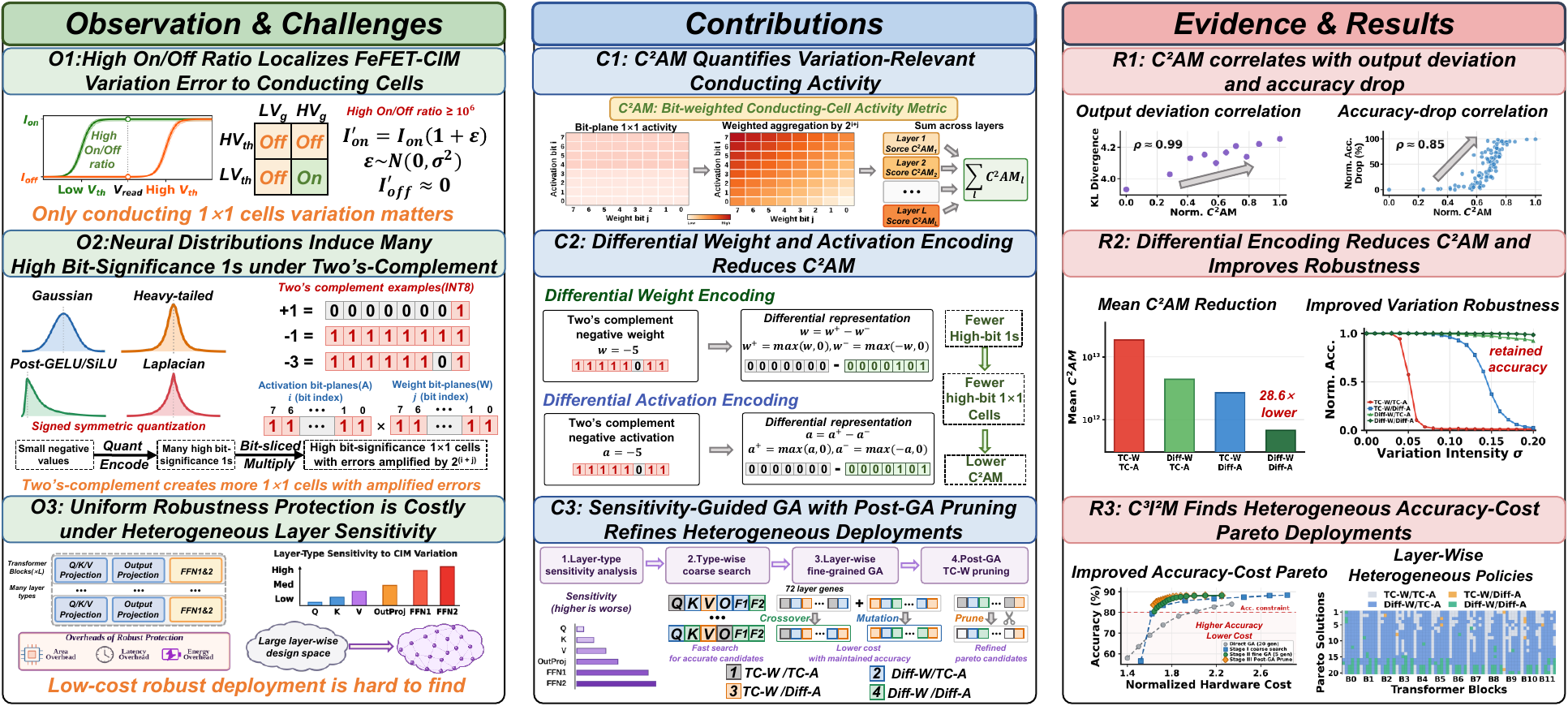}
\caption{Overview of CAFE. CAFE exploits the conducting-cell-dominated variation behavior of FeFET-based CIM and mitigates two's-complement sign-extension amplification through conducting-cell-aware adaptive differential encoding. It uses C\(^2\)AM, a bit-significance-weighted conducting-cell activity metric, to guide differential weight and activation encoding and adaptive layer-wise deployment.}
\label{fig:cafe_overview}
\end{figure*}

Our key observation is that FeFET-CIM variation is not uniformly distributed across all stored bits or cells. Due to the extremely high on/off ratio of FeFETs, off-state leakage is negligible, and variation-induced computation error is dominated by conducting on-state cells. This conducting-cell-dominated behavior becomes especially problematic under two's-complement signed representation: small negative weights or activations are numerically small, but sign extension maps them to many high-significance ones, creating variation-sensitive conducting events whose errors are amplified by shift-and-add reconstruction. Fig.~\ref{fig:cafe_overview} summarizes the three observations that motivate CAFE and the corresponding design components: FeFET variation is centered on conducting cells, two's-complement sign extension induces excessive high-significance conducting activity, and layer sensitivity differs across layer types. Together, these observations make signed representation a device-variation problem in bit-sliced FeFET CIM.
%%%%%%%%%%%%%%%%%%%%%%%%%%%%%%%%%
%%%%%%%%%%%%%%%%%%%%%%%%%%%%%%%%%

Differential encoding has long been used in analog CIM for signed arithmetic. In contrast, bit-sliced CIM commonly relies on two's-complement or offset-style encodings because they align naturally with bit-plane multiplication and shift-and-add reconstruction, whereas differential encoding incurs additional hardware cost. As a result, signed representation choices in bit-sliced CIM are typically evaluated for arithmetic support, representation cost, or numerical error. This view overlooks the FeFET-specific conducting-cell asymmetry: under high on/off ratio devices, signed encodings also determine where variation-sensitive conducting activity appears across bit slices.

To address this gap, we propose \textbf{CAFE}, a \textbf{c}onducting-cell-aware \textbf{a}daptive differential encoding framework for robust DNN inference in \textbf{Fe}FET-based CIM. CAFE treats signed representation not as a fixed arithmetic convention, but as a hardware design variable for FeFET variation robustness. It introduces \textbf{C\(^2\)AM}, a bit-significance-weighted conducting-cell activity metric, to estimate the robustness impact of signed encoding choices. Guided by C\(^2\)AM, CAFE selectively replaces two's-complement weight and activation encodings with differential encodings to suppress sign-extension-induced conducting activity. For deployment, CAFE combines layer-type sensitivity analysis, type-wise encoding enumeration, layer-wise NSGA-II search, and post-GA pruning to balance robustness and hardware cost.
This paper makes the following contributions:
\begin{itemize}
    \item We identify two's-complement sign extension as a key amplifier of FeFET conducting-cell variation and propose C\(^2\)AM, a bit-significance-weighted conducting-cell activity metric for predicting variation-induced robustness degradation.
    \item We propose CAFE, a conducting-cell-aware adaptive differential encoding framework for bit-sliced FeFET CIM, covering differential weight and activation encoding, layer-wise adaptive deployment, and post-GA pruning.
    \item We design a sensitivity-guided deployment flow that combines layer-type analysis, type-wise enumeration, and layer-wise NSGA-II search to allocate differential protection under hardware-cost constraints.
    \item We validate CAFE across vision and language workloads, showing up to \(28.6\times\) C\(^2\)AM reduction and accuracy gains over two's-complement CIM by more than 87 percentage points, while approaching full-differential robustness at \(1.86\)--\(2.78\times\) normalized cost.
\end{itemize}

\section{Background and Related Work}
\label{sec:background_related}

\subsection{FeFET Device}
\label{subsec:fefet}

\begin{figure}[!t]
    \centering
    \includegraphics[width=\columnwidth]{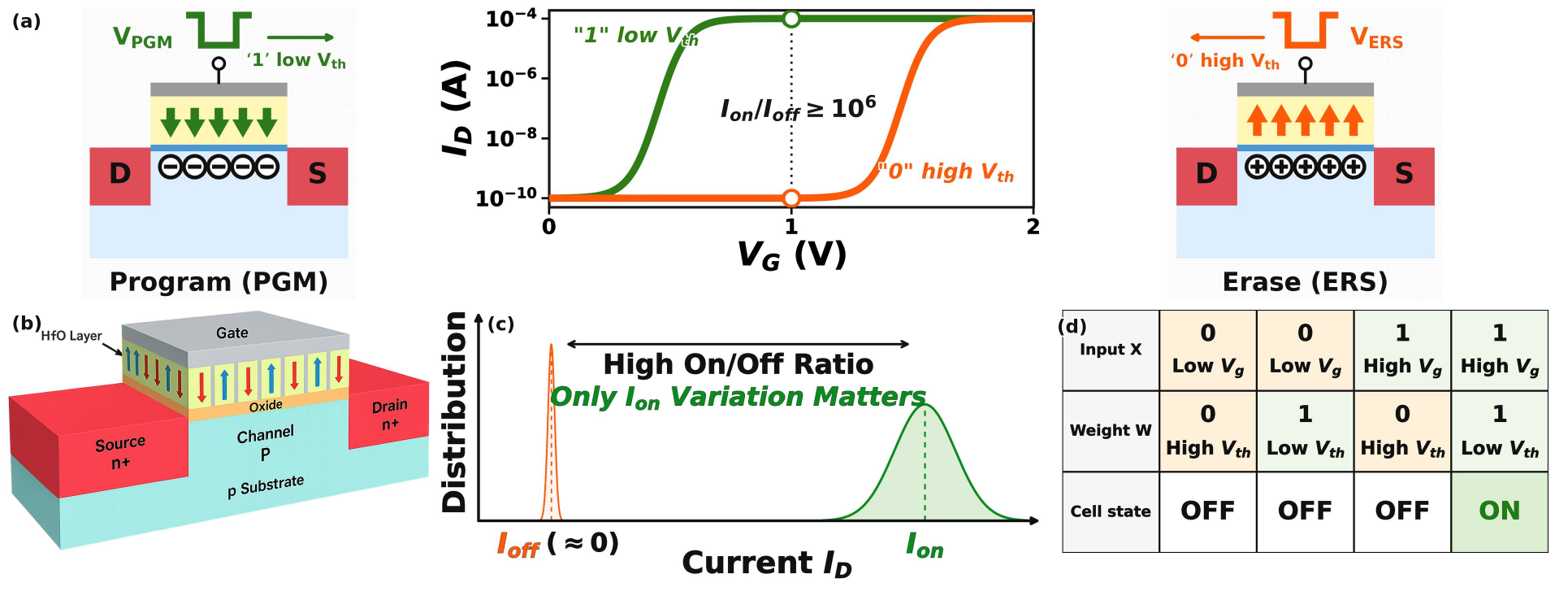}
    \caption{FeFET states, on- and off-current distributions, and cell read behavior.}
    \label{fig:fefet_background}
\end{figure}

An FeFET integrates a ferroelectric layer into the MOSFET gate stack, as shown in Fig.~\ref{fig:fefet_background}(b). Program and erase pulses switch polarization and shift the threshold voltage, creating two nonvolatile storage states with distinct threshold voltages. During read, a gate bias between these states produces a large \(I_{\mathrm{on}}\) for stored ``1'' and a small \(I_{\mathrm{off}}\) for stored ``0'' (Fig.~\ref{fig:fefet_background}(a)).

Because \(I_{\mathrm{on}}\) and \(I_{\mathrm{off}}\) differ by orders of magnitude (Fig.~\ref{fig:fefet_background}(c)), off-state leakage is negligible and variation is dominated by the on-state read current. A cell conducts only when the read gate is high and the device stores ``1'' (Fig.~\ref{fig:fefet_background}(d)).

\subsection{Bit-Sliced CIM and Signed Encoding}
\label{subsec:cim}

\begin{figure}[!t]
    \centering
    \includegraphics[width=\columnwidth]{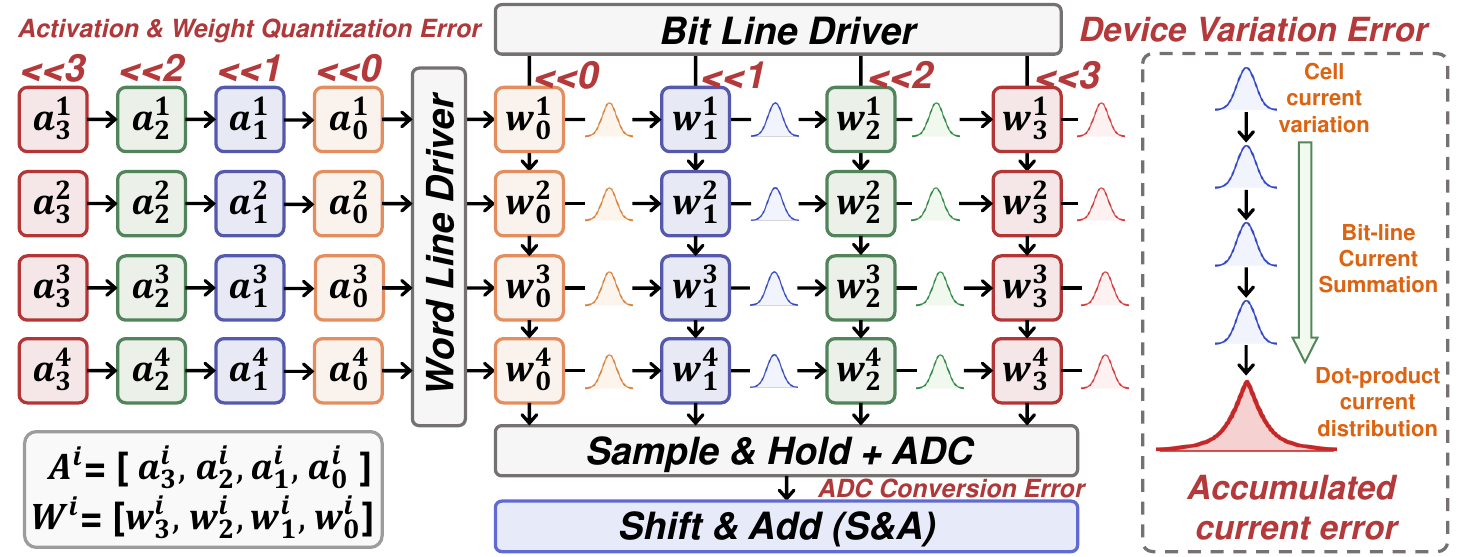}
    \caption{Bit-sliced CIM flow and considered error sources.}
    \label{fig:bitsliced_cim}
\end{figure}

Fig.~\ref{fig:bitsliced_cim} illustrates the bit-serial-input and bit-parallel-weight CIM flow considered in this work. A quantized dot product is decomposed into activation and weight bit-plane products; activation bits are streamed on word lines, weight bits are stored across array columns, and each bit pair is accumulated on bit lines.

Combined with the conducting-cell-dominated FeFET behavior above, shift-and-add reconstruction makes robustness depend on where conducting activity appears across bit planes. Signed computation in bit-sliced CIM commonly uses two's-complement, offset-binary, or canonical signed digit (CSD)-based encodings that align naturally with bit-plane reconstruction~\cite{xiao2023radix4_csd_imac}. Differential encoding has also been used in analog CIM for signed arithmetic~\cite{shafiee2016isaac,chi2016prime}, but uniform differential deployment in bit-sliced CIM incurs extra storage, array, and peripheral cost. CAFE revisits differential encoding as a robustness mechanism: by decomposing signed values into positive and negative components, it removes sign extension and suppresses high-significance conducting-cell activity under FeFET variation.

\subsection{Robustness and Adaptive Deployment}

Prior CIM robustness techniques use hardware compensation, calibration, error correction, or noise-aware training. While useful, they often require specialized hardware support or retraining, still leaving the representation-dependent error structure of bit-sliced FeFET CIM unaddressed. These techniques are largely orthogonal to CAFE, which reshapes signed bit-plane activity before CIM execution. Efficiency methods such as mixed-precision mapping~\cite{wang2019haq} and pruning~\cite{han2015learning} exploit layer-wise sensitivity or sparsity. CAFE instead keeps quantized precision fixed, uses C\(^2\)AM to quantify bit-significance-weighted conducting-cell activity, and applies pruning as a robustness refinement rather than sparse execution.

\section{Methodology}
\label{sec:methodology}

% CAFE is designed around a device- and representation-aware view of FeFET-CIM variation. Instead of treating variation as uniform noise, CAFE focuses on the conducting cells that dominate output deviation under the high on/off ratio of FeFETs. This section first formulates the conducting-cell-dominated variation behavior in bit-sliced FeFET CIM, then defines C2AM as a bit-significance-weighted conducting-cell activity metric. Based on C2AM, we introduce differential weight and activation encoding to suppress two's-complement sign-extension amplification. Finally, we present a sensitivity-guided adaptive deployment procedure that selects layer-wise encoding policies with favorable performance-cost trade-offs.

\subsection{Conducting-Cell-Dominated Variation}
\label{subsec:variation_model}

We consider one output element of a bit-sliced dot product. Let \(w_{r,b}\in\{0,1\}\) be the \(b\)-th stored weight bit and \(x_{r,k}\in\{0,1\}\) be the \(k\)-th activation bit of input element \(r\). Their product
\begin{equation}
    c_{r,b,k}=w_{r,b}x_{r,k}
    \label{eq:conducting_event}
\end{equation}
indicates whether the corresponding FeFET cell conducts for bit pair \((b,k)\). The bit-plane partial sum is \(p_{b,k}=\sum_r c_{r,b,k}\), and the arithmetic output is reconstructed as
\begin{equation}
    y = \sum_{b,k} s_{b,k} p_{b,k},
    \label{eq:shift_add}
\end{equation}
where \(s_{b,k}\) is the signed reconstruction coefficient determined by the encoding and the weight and activation bit positions. For unsigned bit planes, \(s_{b,k}=2^{b+k}\); for signed encodings, sign bits or emitted signed digits change the sign but not the need to weight errors by reconstruction magnitude.

Due to the large FeFET on/off ratio, off-state cells contribute negligible read current (\(I'_{\mathrm{off}}\approx0\)). We model a conducting on-state cell as
\begin{equation}
    I'_{\mathrm{on}} = I_{\mathrm{on}}(1+\epsilon),
    \qquad
    \mathbb{E}[\epsilon]=0,\quad
    \mathrm{Std}[\epsilon]=\sigma,
    \label{eq:on_variation}
\end{equation}
Here \(\epsilon\) is an effective normalized perturbation that absorbs device-to-device (D2D), cycle-to-cycle (C2C), and read-path on-current variation into the standard deviation \(\sigma\); analog-to-digital converter (ADC) quantization is modeled separately in our simulator. Unless otherwise specified, experiments sample \(\epsilon\) as zero-mean Gaussian with standard deviation \(\sigma\). The current perturbation accumulated by bit pair \((b,k)\) is therefore
\begin{equation}
    \Delta I_{b,k}
    =
    I_{\mathrm{on}}\sum_r c_{r,b,k}\epsilon_{r,b,k}.
    \label{eq:partial_error}
\end{equation}
Ignoring fixed current-to-digital conversion constants, shift-and-add reconstruction maps these bit-plane perturbations to
\begin{equation}
    \Delta y
    =
    \sum_{b,k}
    s_{b,k}
    \Delta I_{b,k}.
    \label{eq:output_error}
\end{equation}
Taking magnitudes gives a representation-dependent error envelope:
\begin{equation}
    |\Delta y|
    \le
    I_{\mathrm{on}}
    \sum_{b,k}|s_{b,k}|
    \sum_r c_{r,b,k}|\epsilon_{r,b,k}|.
    \label{eq:error_bound}
\end{equation}
Thus, the sign of \(s_{b,k}\) determines addition or subtraction, while \(|s_{b,k}|\) determines how strongly a conducting-cell perturbation is amplified. FeFET-CIM variation is therefore concentrated on conducting \(1\times1\) events, especially those appearing in high-significance bit slices.

%%%%%%%%%%%%%%%%%%%%%%%%%%%%%%%%%%%%%%%%%%%
\begin{figure*}[!t]
\centering
\includegraphics[width=\textwidth]{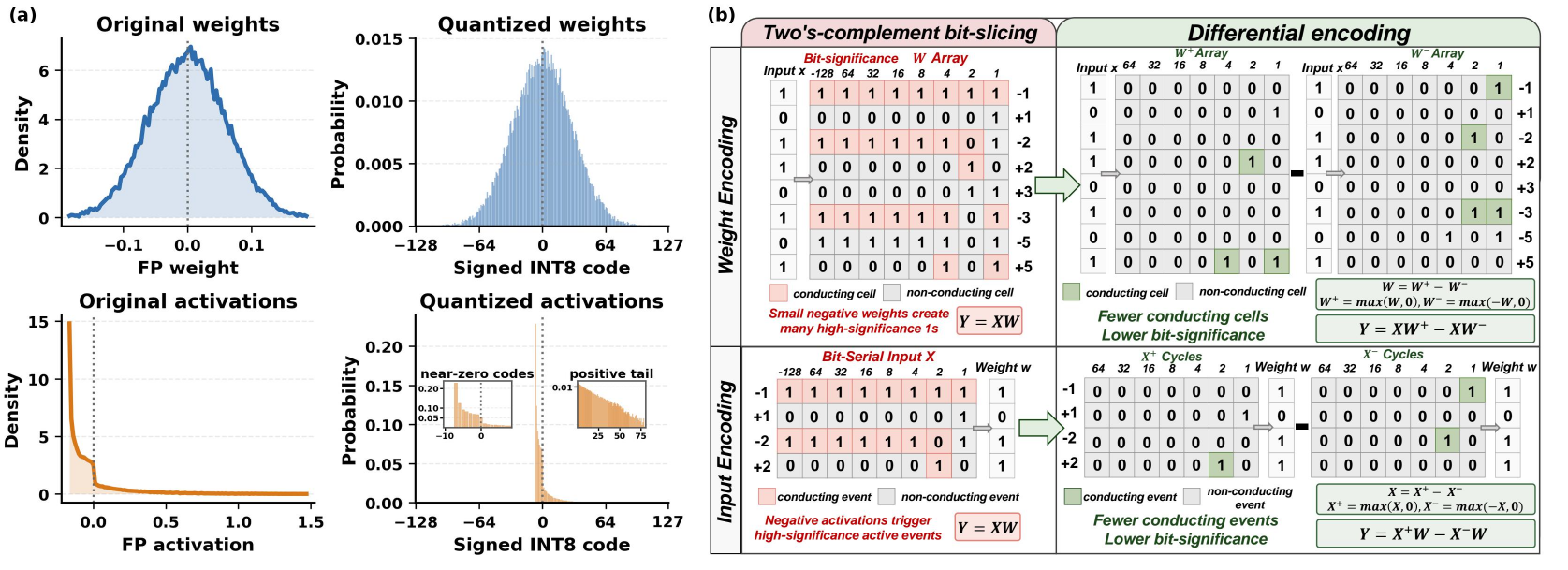}
\caption{Sign-extension amplification and differential encoding. (a) Near-zero negative quantized values create high-significance ones under two's-complement bit-slicing. (b) Differential weight and activation encoding removes sign extension with positive and negative unsigned components.}
\label{fig:differential_encoding}
\end{figure*}
%%%%%%%%%%%%%%%%%%%%%%%%%%%%%%%%%%%%%%%%%%%

\subsection{C\(^2\)AM: Conducting-Cell Activity Metric}
\label{subsec:ccam}

Based on Eq.~\eqref{eq:error_bound}, we define C\(^2\)AM as a bit-significance-weighted count of conducting activity. For layer \(l\), the number of conducting events at bit pair \((b,k)\) is
\begin{equation}
    N_{1\times1}^{(l)}(b,k)
    =
    \sum_r c_{r,b,k}^{(l)},
    \label{eq:conducting_count}
\end{equation}
where the count is accumulated over evaluated samples or tokens, output channels, and reduction dimensions. The model-level metric is
\begin{equation}
    \mathrm{C}^{2}\mathrm{AM}
    =
    \sum_l
    \sum_{b,k}
    |s_{b,k}^{(l)}|
    N_{1\times1}^{(l)}(b,k).
    \label{eq:c2am}
\end{equation}
For Gaussian effective variation, \(\mathbb{E}|\epsilon|=\sigma\sqrt{2/\pi}\), so C\(^2\)AM is proportional to the representation-dependent part of the expected error envelope in Eq.~\eqref{eq:error_bound}. It is computed directly from encoded bit planes and does not require Monte Carlo variation simulation, making it suitable for rapid comparison among encoding choices.

C\(^2\)AM differs from generic activity counting in two aspects. First, it counts only conducting \(1\times1\) events, matching the FeFET-specific observation that off-state leakage is negligible. Second, it weights each event by bit significance, reflecting shift-and-add amplification. As a result, encodings that represent the same quantized tensor values can still have different C\(^2\)AM values when they induce different high-significance conducting patterns.

C\(^2\)AM is validated at both the sample-output and end-to-end accuracy levels in Sec.~\ref{sec:experiments}.

\subsection{C\(^2\)AM-Guided Differential Encoding}
\label{subsec:differential_encoding}
%%%%%%%%%%%%%%%%%%%%%%%%%%%%%%%%%%%%%%%%%%%

A major representation-dependent source of high-significance conducting activity in signed bit-sliced CIM is two's-complement sign extension. For a \(B\)-bit signed value \(z\), two's-complement representation computes
\begin{equation}
    z =
    -2^{B-1}z_{B-1}
    +
    \sum_{b=0}^{B-2}
    2^b z_b ,
    \label{eq:twos_complement}
\end{equation}
where \(z_{B-1}\) is the sign bit. Sign extension makes negative codes carry leading ones in high-significance bit planes. Because quantized DNN weights are often concentrated near zero and signed activations can also contain near-zero negative codes, these patterns frequently appear in practical layers, as illustrated in Fig.~\ref{fig:differential_encoding}(a). In bit-sliced FeFET CIM, sign-extension ones are physical stored or streamed bits; when paired with one bits from the other operand, they increase high-significance conducting \(1\times1\) events whose variations are amplified by shift-and-add reconstruction. The represented integer is fixed, but the physical conducting pattern changes.

CAFE uses differential encoding to change the physical bit representation while preserving the signed quantized tensor. For weights,
\begin{IEEEeqnarray}{c}
W = W^{+}-W^{-},\, W^{+}=\max(W,0),\, W^{-}=\max(-W,0)\IEEEeqnarraynumspace\label{eq:diff_weight}
\end{IEEEeqnarray}
where \(W^+\) and \(W^-\) are unsigned magnitude tensors. A weight-differential layer computes
\begin{equation}
    Y = XW = XW^+ - XW^- .
    \label{eq:diff_weight_output}
\end{equation}
Similarly, activation differential encoding decomposes
\begin{IEEEeqnarray}{c}
X = X^{+}-X^{-},\, X^{+}=\max(X,0),\, X^{-}=\max(-X,0)\IEEEeqnarraynumspace\label{eq:diff_activation}
\end{IEEEeqnarray}
and computes
\begin{equation}
    Y = XW = X^+W - X^-W .
    \label{eq:diff_activation_output}
\end{equation}
When both operands are differentially encoded, the multiplication becomes
\begin{equation}
    Y = X^+W^+ - X^+W^- - X^-W^+ + X^-W^- .
    \label{eq:full_diff_output}
\end{equation}
The signs are applied during digital recombination, while CIM arrays store and multiply only unsigned magnitude bit planes. Thus the clean quantized arithmetic is unchanged, but sign-extension ones are removed from CIM execution, as shown in Fig.~\ref{fig:differential_encoding}(b). This representation change reduces \(N_{1\times1}^{(l)}(b,k)\) on high-significance bit pairs in Eq.~\eqref{eq:error_bound}, thereby lowering C\(^2\)AM without retraining or changing quantization scales.

CAFE considers four encoding configurations for each layer: \(\mathrm{TC\text{-}W/TC\text{-}A}\), \(\mathrm{Diff\text{-}W/TC\text{-}A}\), \(\mathrm{TC\text{-}W/Diff\text{-}A}\), and \(\mathrm{Diff\text{-}W/Diff\text{-}A}\), denoted as \(\mathcal{E}\), where \(\mathrm{TC}\) denotes two's-complement encoding and \(\mathrm{Diff}\) denotes differential encoding. \(\mathrm{TC\text{-}W/TC\text{-}A}\) is the lowest-cost baseline, while \(\mathrm{Diff\text{-}W/Diff\text{-}A}\) provides the strongest sign-extension removal but incurs the highest branch, storage, and bit-serial overhead. Therefore, differential encoding should be applied selectively rather than uniformly across the whole model.
%%%%%%%%%%%%%%%%%%%%%%%%%%%%%%%%%%%%%%%%%%%
\begin{figure*}[!t]
\centering
\includegraphics[width=\textwidth]{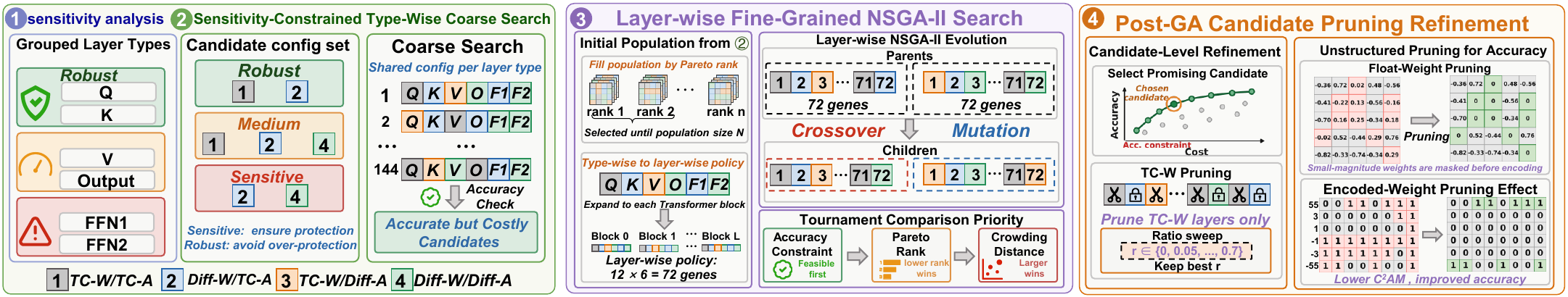}
\caption{CAFE adaptive deployment flow: layer-type sensitivity guides type-wise enumeration, layer-wise NSGA-II search, and post-GA pruning for robustness-cost optimization.}
\label{fig:adaptive_deployment}
\end{figure*}
%%%%%%%%%%%%%%%%%%%%%%%%%%%%%%%%%%%%%%%%%%%

\subsection{Sensitivity-Guided Adaptive Deployment}
\label{subsec:adaptive_deployment}

Full differential encoding reduces C\(^2\)AM but incurs high cost when applied uniformly. To balance robustness and cost, CAFE combines layer-type sensitivity, type-wise coarse search, layer-wise non-dominated sorting genetic algorithm II (NSGA-II)~\cite{deb2002nsga2}, and post-GA pruning to select adaptive encoding schemes from \(\mathcal{E}\), as shown in Fig.~\ref{fig:adaptive_deployment}.

We use an area-latency cost proxy, normalized to the all-\(\mathrm{TC\text{-}W/TC\text{-}A}\) baseline, as a search-ranking objective rather than an absolute layout estimator. Differential weights require two unsigned magnitude branches with \(2(B_w-1)\) bit planes, while differential activations require \(2(B_a-1)\) bit-serial cycles. For layer \(\ell\), let \(c^W_\ell\in\{1,2(B_w-1)/B_w\}\), \(c^A_\ell\in\{1,2(B_a-1)/B_a\}\), and let \(\lambda_\ell\) be proportional to layer size. The normalized cost is
\begin{equation}
    C_{\mathrm{norm}}
    =
    \frac{\sum_{\ell}\lambda_{\ell} c^W_{\ell}c^A_{\ell}}
         {\sum_{\ell}\lambda_{\ell}}.
    \label{eq:norm_hw_cost}
\end{equation}
This proxy charges explicit branch and cycle overhead but does not credit possible peripheral savings from reduced bit-line current range or relaxed ADC precision, making the reported cost conservative for differential policies.

\subsubsection{Layer-Type Sensitivity Analysis}
CAFE profiles Transformer layer types by injecting FeFET-CIM variation into one functional role at a time, such as query, key, value, output, or feed-forward network (FFN) projections. Types causing larger accuracy degradation receive stronger candidate protections, while robust types retain lower-cost options.

\subsubsection{Sensitivity-Constrained Type-Wise Coarse Search}
Given \(\mathcal{E}\), exhaustive layer-wise search is costly. CAFE therefore first enumerates type-wise policies in which all layers of the same type share one encoding. The allowed encoding subset for each type is constrained by its sensitivity group: robust types can choose low-cost encodings, while sensitive types must include stronger protection. CAFE evaluates robust accuracy and cost for these coarse policies and keeps accurate candidates as Pareto-ranked seeds.

\subsubsection{Layer-Wise Fine-Grained NSGA-II Search}
To recover flexibility beyond shared type-wise assignments, CAFE expands promising type-wise policies into layer-wise chromosomes, where each gene \(e_l\in\mathcal{E}\) specifies one CIM layer's encoding. For a Transformer with \(B_{\mathrm{blk}}\) blocks and six profiled layer types per block, the chromosome has \(6B_{\mathrm{blk}}\) genes. NSGA-II maximizes validation accuracy under FeFET-CIM variation while minimizing \(C_{\mathrm{norm}}\). The population is initialized from Pareto-ranked type-wise candidates and evolved by crossover and mutation, with tournament selection prioritizing feasibility, lower Pareto rank, and larger crowding distance.
% To improve efficiency, CAFE initializes NSGA-II from Pareto-ranked type-wise candidates instead of random policies, producing diverse accuracy-cost Pareto solutions.

\subsubsection{Post-GA Pruning Refinement}
After NSGA-II, CAFE applies post-GA pruning to selected Pareto candidates. This step targets harmful small-magnitude weights that may become small negative two's-complement codes and introduce high-significance conducting activity, rather than performing general model compression. Since differential encoding already removes sign extension, pruning is mainly applied to layers using two's-complement weight encoding.

For each selected candidate, CAFE sweeps the pruning ratio \(r\in\{0,0.05,\ldots,0.7\}\). Small-magnitude floating-point weights are masked before quantization and encoding. The pruned candidates are then evaluated under FeFET-CIM variation, and CAFE keeps the one with the best accuracy-cost trade-off.

\section{Experiments}
\label{sec:experiments}

\subsection{Experimental Setup}
\label{subsec:setup}

We evaluate CAFE with a behavioral bit-sliced CIM simulator adapted from NeuroSim~\cite{lu2021neurosim}, extended to support signed Transformer bit-plane encodings and the conducting-cell-level FeFET error flow in Fig.~\ref{fig:bitsliced_cim}, including bit-plane decomposition, on-current variation, bit-line accumulation, ADC quantization, and shift-and-add reconstruction. The main workload is Vision Transformer (ViT)-Tiny~\cite{dosovitskiy2021vit} on CIFAR-100~\cite{krizhevsky2009cifar}, with additional evaluations on ViT-Base on CIFAR-100, ViT-Tiny on ImageNet, and Pythia~\cite{biderman2023pythia} on WikiText-103.

Motivated by reported FeFET D2D and C2C variation~\cite{manna2024fefet_variation}, we sweep an effective on-current perturbation strength from \(\sigma=0\) to \(0.20\) with step \(0.01\), using \(\sigma=0\) as the clean reference. Unless specified, weights use 8-bit per-channel quantization, activations use 8-bit per-token quantization, and the CIM simulator uses a 6-bit ADC and 64 parallel read rows; variation results are averaged over 20 seeds from 2026 to 2045. Baselines include TC-W/TC-A, Diff-W/TC-A, TC-W/Diff-A, Diff-W/Diff-A, offset-binary, and CSD encoding~\cite{xiao2023radix4_csd_imac}. Deployment experiments report normalized cost using Eq.~\eqref{eq:norm_hw_cost}.

\subsection{Experimental Results}
\begin{figure}[!t]
\centering
\includegraphics[width=\columnwidth]{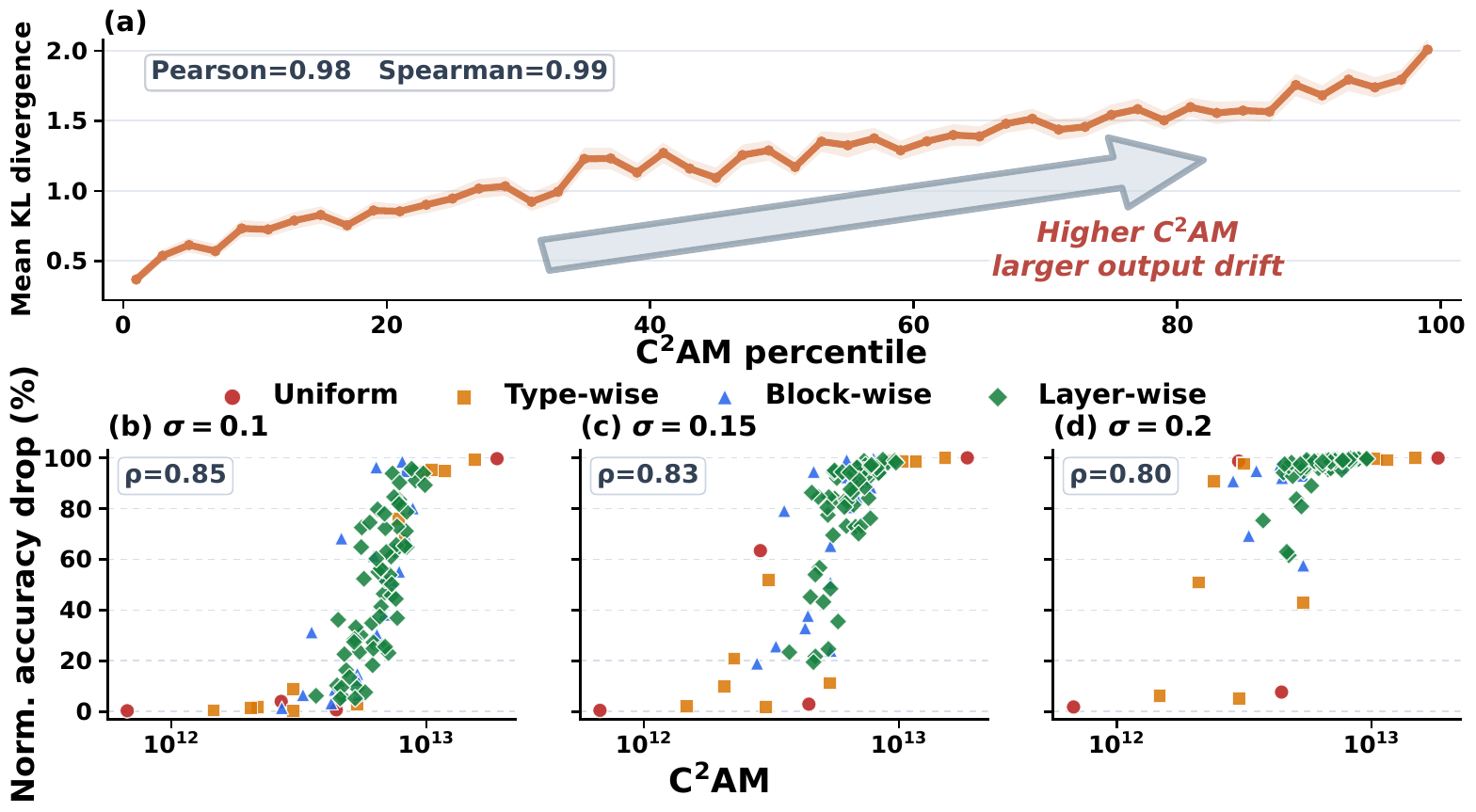}
\caption{Multi-level validation of C\(^2\)AM: (a) sample-level output deviation and (b)--(d) policy-level accuracy drop.}
\label{fig:c2am_validation}
\end{figure}

% \textbf{C2AM predicts robustness degradation across deployment granularities.}
Fig.~\ref{fig:c2am_validation} validates C\(^2\)AM at both sample and policy levels. Fig.~\ref{fig:c2am_validation}(a) shows that per-sample C\(^2\)AM tracks the Kullback-Leibler (KL) divergence between noisy and clean outputs under the two's-complement baseline, consistent with the error envelope in Eq.~\eqref{eq:error_bound}. Fig.~\ref{fig:c2am_validation}(b)--(d) further evaluates diagnostic policies spanning uniform, type-wise, block-wise, and layer-wise encodings, showing that policies with larger C\(^2\)AM consistently suffer larger normalized accuracy degradation under different FeFET variation strengths. The monotonic correlation remains strong as \(\sigma\) increases, with Spearman coefficients of \(\rho=0.85\), \(\rho=0.83\), and \(\rho=0.80\) for \(\sigma=0.1\), \(0.15\), and \(0.2\), respectively. This multi-level validation supports using C\(^2\)AM as a low-cost robustness indicator during encoding search.

\begin{figure}[!t]
\centering
\includegraphics[width=\columnwidth]{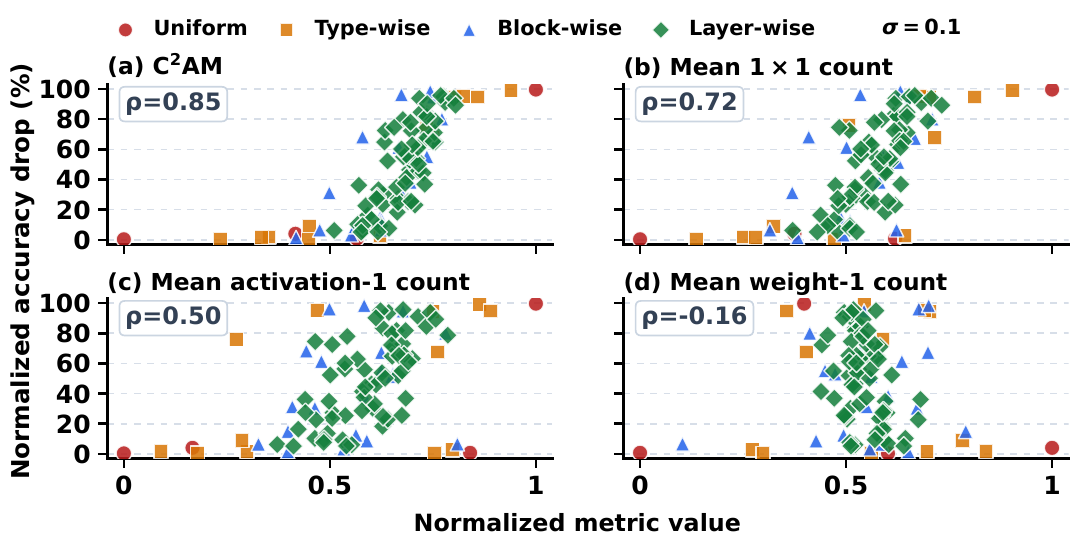}
\caption{Comparison between C\(^2\)AM and alternative activity metrics.}
\label{fig:metric_comparison}
\end{figure}

% \textbf{Bit-significance weighting is essential for predicting accuracy degradation.}
Fig.~\ref{fig:metric_comparison} compares C\(^2\)AM with normalized count-based activity metrics at \(\sigma=0.1\). C\(^2\)AM achieves the strongest monotonic correlation with accuracy drop, while the mean \(1\times1\) count is weaker because it ignores bit significance, and activation-only or weight-only counts are less predictive. This confirms that FeFET-CIM robustness depends on both conducting-cell activity and shift-and-add significance.

\begin{figure}[!t]
\centering
\includegraphics[width=\columnwidth]{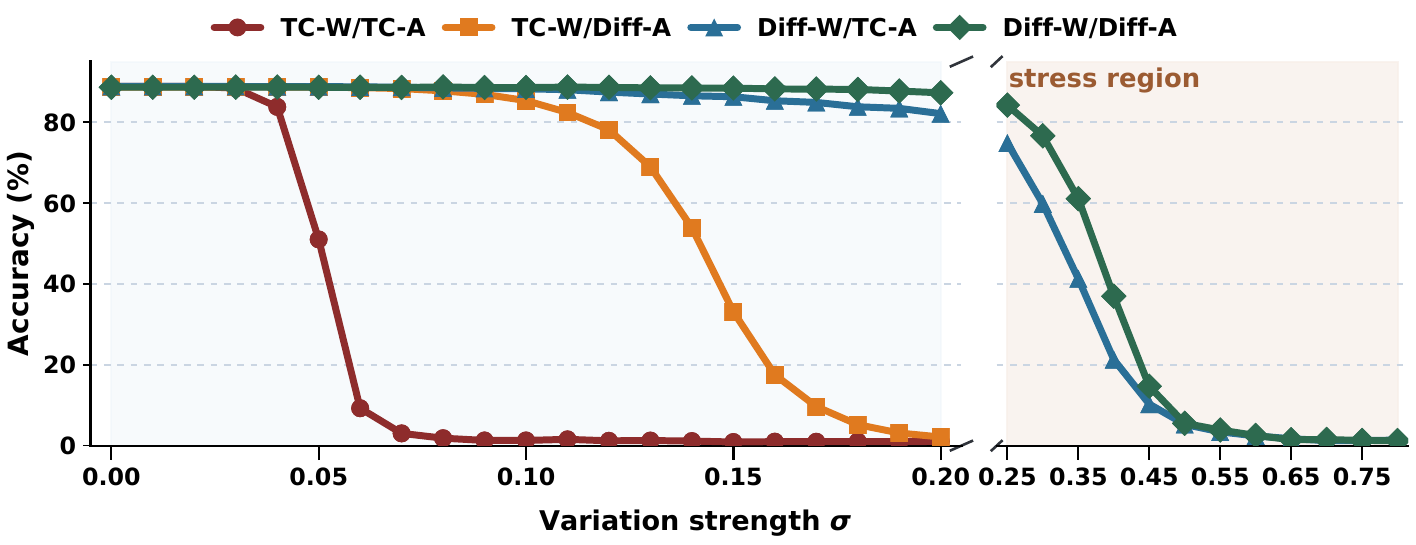}
\caption{Accuracy of different uniform encoding policies under FeFET variation from the default sweep to an extended stress setting.}
\label{fig:uniform_encoding_variation}
\end{figure}

% \textbf{Differential encoding substantially improves variation robustness.}
Fig.~\ref{fig:uniform_encoding_variation} and Table~\ref{tab:encoding_and_deployment_summary}(a) evaluate uniform encoding policies under the default FeFET variation setting. The figure shows the accuracy trend across the default and extended stress ranges, while the table reports mean accuracy, normalized C\(^2\)AM, and hardware cost.

\begin{table*}[!t]
\centering
\caption{Summary of encoding baselines and CAFE deployments: (a) uniform encoding accuracy, normalized C\(^2\)AM, and hardware cost under Gaussian FeFET variation; (b) robustness under truncated-normal and lognormal variation; and (c) representative adaptive deployments selected at different variation strengths.}
\label{tab:encoding_and_deployment_summary}
\vspace{-0.7em}

\definecolor{EncBest}{HTML}{FCE8D9}
\definecolor{DeployBest}{HTML}{DFF0E6}
\definecolor{HeadFill}{HTML}{EEF1F6}
\definecolor{SubHeadFill}{HTML}{F7F8FB}
\begingroup
\fontsize{7.0pt}{7.3pt}\selectfont
\setlength{\tabcolsep}{1.05pt}
\renewcommand{\arraystretch}{0.82}

% -------------------- Uniform encoding table --------------------
\begin{tabular}{@{}>{\raggedright\arraybackslash}p{0.168\textwidth}
                >{\centering\arraybackslash}p{0.069\textwidth}
                >{\centering\arraybackslash}p{0.134\textwidth}
                >{\centering\arraybackslash}p{0.134\textwidth}
                >{\centering\arraybackslash}p{0.134\textwidth}
                >{\centering\arraybackslash}p{0.134\textwidth}
                >{\centering\arraybackslash}p{0.124\textwidth}
                >{\centering\arraybackslash}p{0.066\textwidth}@{}}
\toprule
\rowcolor{HeadFill}
\textbf{(a) Encoding} &
\(\boldsymbol{\sigma=0}\) &
\(\boldsymbol{0.05}\) &
\(\boldsymbol{0.10}\) &
\(\boldsymbol{0.15}\) &
\(\boldsymbol{0.20}\) &
\textbf{Norm. C\(^2\)AM} &
\textbf{Cost} \\
\midrule
TC-W/TC-A       & 88.88 & 50.65\(\pm\)2.35 & 1.39\(\pm\)0.12  & 1.13\(\pm\)0.13  & 1.11\(\pm\)0.07  & 1.000 & 1.00 \\
Offset-W/TC-A   & 88.88 & 44.93\(\pm\)1.64 & 1.37\(\pm\)0.09  & 1.14\(\pm\)0.10  & 1.07\(\pm\)0.10  & 1.016 & 1.00 \\
CSD-W/TC-A      & 88.88 & 88.86\(\pm\)0.07 & 87.73\(\pm\)0.20 & 81.55\(\pm\)0.53 & 62.16\(\pm\)2.51 & 0.301 & 2.00 \\
TC-W/Diff-A     & 88.66 & 88.62\(\pm\)0.08 & 85.40\(\pm\)0.33 & 36.67\(\pm\)3.77 & 2.28\(\pm\)0.38  & 0.140 & 1.75 \\
Diff-W/TC-A     & 88.88 & 88.86\(\pm\)0.06 & 88.29\(\pm\)0.13 & 86.18\(\pm\)0.19 & 81.96\(\pm\)0.32 & 0.231 & 1.75 \\
\rowcolor{EncBest}
\textbf{Diff-W/Diff-A}   & \textbf{88.66} & \textbf{88.67\(\pm\)0.04} & \textbf{88.65\(\pm\)0.07} & \textbf{88.40\(\pm\)0.09} & \textbf{87.21\(\pm\)0.18} & \textbf{0.035} & \textbf{3.06} \\
\bottomrule
\end{tabular}%

\vspace{0.45pt}

% -------------------- Bottom two tables --------------------
\noindent
\makebox[\textwidth][c]{%
\begin{minipage}[t]{0.620\textwidth}
\centering
\vspace{0pt}
\setlength{\tabcolsep}{0.85pt}
\renewcommand{\arraystretch}{0.82}
\begin{tabular}{@{}>{\raggedright\arraybackslash}p{0.220\linewidth}
                >{\centering\arraybackslash}p{0.092\linewidth}
                >{\centering\arraybackslash}p{0.092\linewidth}
                >{\centering\arraybackslash}p{0.092\linewidth}
                >{\centering\arraybackslash}p{0.092\linewidth}
                >{\centering\arraybackslash}p{0.092\linewidth}
                >{\centering\arraybackslash}p{0.092\linewidth}
                >{\centering\arraybackslash}p{0.092\linewidth}
                >{\centering\arraybackslash}p{0.092\linewidth}@{}}
\toprule
\rowcolor{HeadFill}
\textbf{(b)} & \multicolumn{4}{c}{\textbf{Truncated normal}} & \multicolumn{4}{c}{\textbf{Lognormal}} \\
\cmidrule(lr){2-5}\cmidrule(lr){6-9}
\rowcolor{SubHeadFill}
\textbf{Encoding} & \(\boldsymbol{0.05}\) & \(\boldsymbol{0.10}\) & \(\boldsymbol{0.15}\) & \(\boldsymbol{0.20}\)
& \(\boldsymbol{0.05}\) & \(\boldsymbol{0.10}\) & \(\boldsymbol{0.15}\) & \(\boldsymbol{0.20}\) \\
\midrule
TC-W/TC-A      & 77.0 & 1.5  & 1.3  & 1.1  & 50.1 & 1.4  & 1.0  & 0.9 \\
TC-W/Diff-A    & 88.6 & 87.6 & 72.2 & 9.7  & 88.7 & 85.2 & 29.0 & 2.5 \\
Diff-W/TC-A    & 88.9 & 88.7 & 87.0 & 84.0 & 88.8 & 88.2 & 86.1 & 82.3 \\
\rowcolor{EncBest}
\textbf{Diff-W/Diff-A}  & \textbf{88.7} & \textbf{88.8} & \textbf{88.8} & \textbf{88.4} & \textbf{88.7} & \textbf{88.5} & \textbf{88.3} & \textbf{87.2} \\
\bottomrule
\end{tabular}%
\end{minipage}%
\hspace{0.004\textwidth}%
\begin{minipage}[t]{0.368\textwidth}
\centering
\vspace{0pt}
\setlength{\tabcolsep}{0.85pt}
\renewcommand{\arraystretch}{1.54}
\begin{tabular}{@{}>{\centering\arraybackslash}p{0.220\linewidth}
                >{\centering\arraybackslash}p{0.125\linewidth}
                >{\centering\arraybackslash}p{0.145\linewidth}
                >{\centering\arraybackslash}p{0.320\linewidth}
                >{\centering\arraybackslash}p{0.145\linewidth}@{}}
\toprule
\rowcolor{HeadFill}
\textbf{(c) Method} & \(\boldsymbol{\sigma}\) & \textbf{Acc.} & \textbf{Norm. C\(^2\)AM} & \textbf{Cost} \\
\midrule
\rowcolor{DeployBest}
 & \textbf{0.10} & \textbf{88.59} & \textbf{0.2166} & \textbf{1.859} \\
\rowcolor{DeployBest}
 & \textbf{0.15} & \textbf{88.10} & \textbf{0.1094} & \textbf{2.556} \\
\rowcolor{DeployBest}
\multirow{-3}{*}{\cellcolor{DeployBest}\textbf{CAFE}} & \textbf{0.20} & \textbf{87.07} & \textbf{0.0755} & \textbf{2.784} \\
\bottomrule
\end{tabular}%
\end{minipage}%
}
\endgroup

\end{table*}

% \textbf{Lower C2AM consistently leads to stronger robustness, but full differential encoding is not always cost-efficient.}
The TC-W/TC-A baseline degrades sharply even at moderate variation, indicating that two's-complement sign extension creates high-significance conducting activity that is highly vulnerable to device perturbation. One-sided differential encoding improves robustness by removing sign-extension activity from either weights or activations, with Diff-W/TC-A remaining more stable in the default sweep. Diff-W/Diff-A maintains near-clean accuracy throughout the default range and shows the largest robustness margin in the extended stress region.

Table~\ref{tab:encoding_and_deployment_summary}(a) further shows that TC-W/TC-A and Offset-W/TC-A have the lowest cost but retain high normalized C\(^2\)AM, so changing the signed representation alone is insufficient. CSD-W/TC-A lowers C\(^2\)AM but still degrades under strong variation and costs more. By directly suppressing sign-extension-induced conducting activity, Diff-W/Diff-A achieves the lowest normalized C\(^2\)AM (0.035, a 28.6\(\times\) reduction) and the best strong-variation accuracy.

Table~\ref{tab:encoding_and_deployment_summary}(b) further evaluates truncated-normal and lognormal perturbations. The same ordering is largely preserved, with Diff-W/Diff-A remaining strongest, indicating that the benefit is not tied to a specific Gaussian error assumption. Table~\ref{tab:encoding_and_deployment_summary}(c) reports representative CAFE deployments for reference; the search process and Pareto trade-offs are analyzed in the adaptive-deployment results.

\begin{figure}[!t]
\centering
\includegraphics[width=\columnwidth]{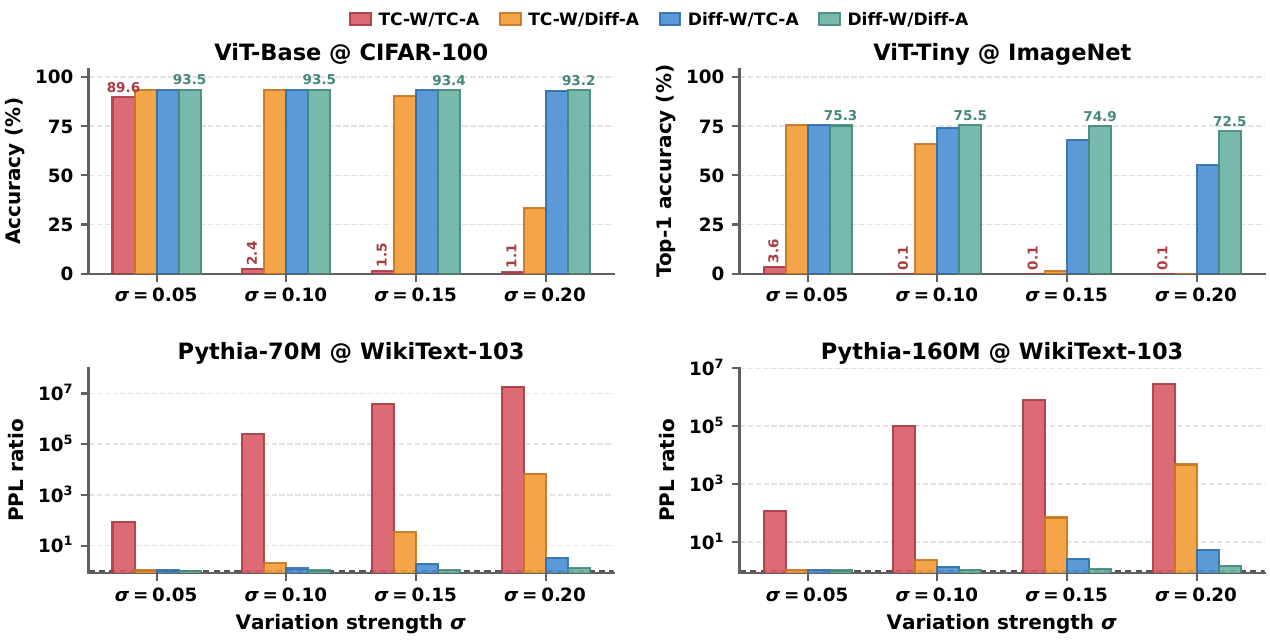}
\caption{Generalization across vision and language workloads.}
\label{fig:generalization_workloads}
\end{figure}

% \textbf{The robustness benefit of differential encoding generalizes across models, datasets, and tasks.}
Fig.~\ref{fig:generalization_workloads} evaluates the four uniform encoding policies beyond the main ViT-Tiny on CIFAR-100 setting, covering ViT-Base on CIFAR-100, ViT-Tiny on ImageNet, and Pythia-70M and Pythia-160M on WikiText-103. Across vision and language workloads, TC-W/TC-A degrades sharply under increasing variation, whereas differential modes preserve accuracy or suppress perplexity growth, with Diff-W/Diff-A consistently strongest. This cross-workload trend shows that the benefit is not tied to one model, dataset, or output metric, but follows from reducing a general representation-induced vulnerability in bit-sliced FeFET CIM.
% These results show that the proposed conductance-aware encoding strategy is not specific to one model scale, dataset, or output metric, but instead addresses a general representation-induced vulnerability in FeFET-based bit-sliced CIM.

\begin{figure}[!t]
\centering
\includegraphics[width=\columnwidth]{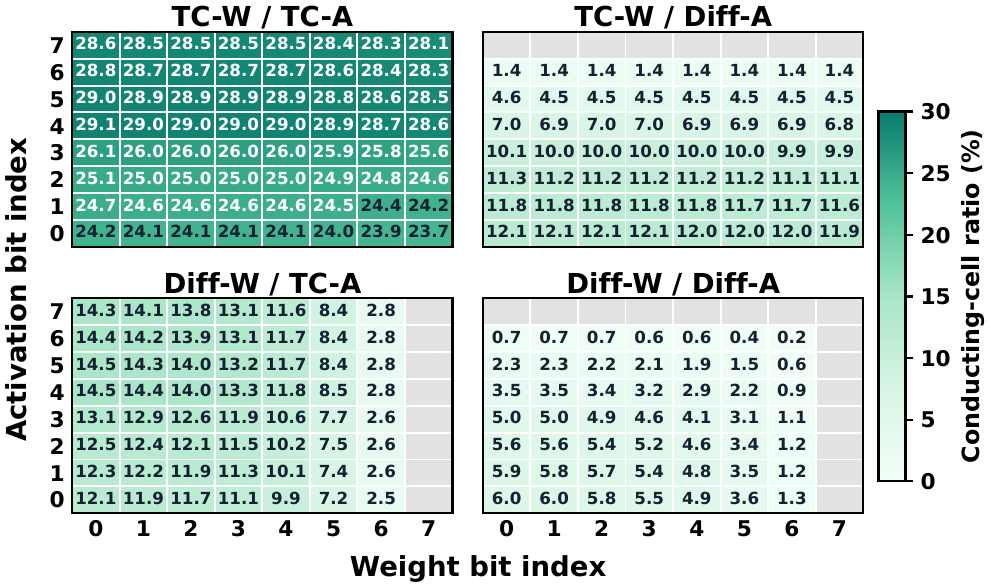}
\caption{Bit-plane \(1\times1\) conducting-event ratio under different encoding policies.}
\label{fig:bitplane_interaction}
\end{figure}
% \textbf{Differential encoding reduces high-significance conducting-cell interactions.}
Fig.~\ref{fig:bitplane_interaction} visualizes the bit-plane \(1\times1\) conducting-event ratio, defined as the fraction of activation-weight bit-pair evaluations that activate conducting FeFET cells. TC-W/TC-A produces dense conducting events across both dimensions, including high-significance regions where variation errors are amplified by shift-and-add reconstruction. Differential encoding suppresses high-significance conducting events along the protected dimension: TC-W/Diff-A reduces activation-side events, Diff-W/TC-A reduces weight-side ones, and Diff-W/Diff-A removes both. As a result, Diff-W/Diff-A reduces the average conducting-event ratio by 8.2\(\times\) overall and by 41\(\times\) in high-significance regions, explaining its lower C\(^2\)AM and stronger robustness.

\begin{figure}[!t]
\centering
\includegraphics[width=\columnwidth]{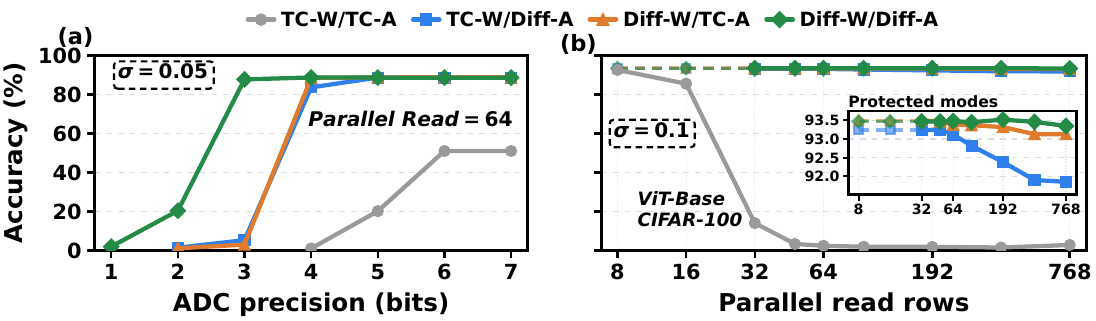}
\caption{Hardware hyperparameter sweeps: (a) ADC precision and (b) parallel read rows.}
\label{fig:hardware_sensitivity}
\end{figure}

% \textbf{Differential encoding improves robustness under hardware-level constraints.}
Fig.~\ref{fig:hardware_sensitivity} sweeps ADC precision and parallel-read rows. The reduced conducting-event ratio in Fig.~\ref{fig:bitplane_interaction} narrows the bit-line current range and lowers ADC requirements: Diff-W/Diff-A reaches 87.8\% with a 3-bit ADC, whereas TC-W/TC-A reaches only 51.0\% even with 6 bits. Under increasing parallel-read rows on ViT-Base on CIFAR-100, fewer active cells let Diff-W/Diff-A remain at 93.4--93.5\% from 32 to 768 rows, while TC-W/TC-A falls to \(\le2.8\%\) beyond 64 rows. These ADC and bit-line tolerance gains are not credited by the normalized cost proxy in Eq.~\eqref{eq:norm_hw_cost}.
% These results indicate that the benefits of conductance-aware encoding are not limited to algorithmic robustness, but also translate into improved tolerance to practical CIM hardware hyperparameters.

\begin{figure}[!t]
\centering
\includegraphics[width=\columnwidth]{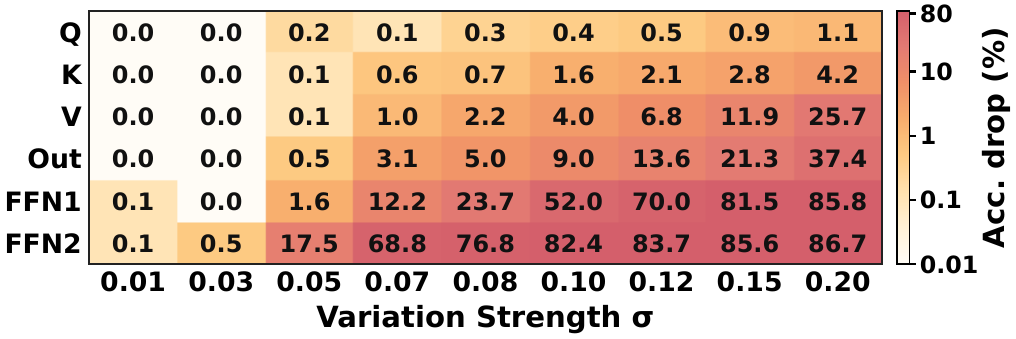}
\caption{Layer-type sensitivity to FeFET-CIM variation.}
\label{fig:layer_type_sensitivity}
\end{figure}

% \textbf{Transformer layer types exhibit strong heterogeneous sensitivity to CIM variation.}
Fig.~\ref{fig:layer_type_sensitivity} profiles layer-type sensitivity by injecting FeFET-CIM variation into one Transformer layer type at a time. Accuracy loss is highly non-uniform: query and key projections are comparatively robust, value and output projections are moderate, and FFN layers, especially FFN2, dominate degradation under strong variation. This non-uniformity motivates type-wise seeding and selective layer-wise protection instead of uniform differential encoding.
% Instead, layer-type sensitivity provides a useful prior for constraining the search space and guiding differential encoding toward the layers where it is most beneficial.

\begin{figure}[!t]
\centering
\includegraphics[width=\columnwidth]{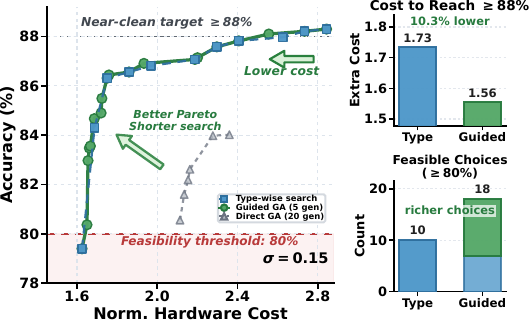}
\caption{Accuracy-cost Pareto comparison.}
\label{fig:pareto_sigma015}
\end{figure}

% \textbf{Sensitivity-guided search finds better Pareto deployments with lower cost and shorter search.}
Fig.~\ref{fig:pareto_sigma015} compares deployment search strategies at \(\sigma=0.15\). Type-wise search quickly finds feasible policies but is limited by shared assignments within each layer type, while direct layer-wise NSGA-II explores a larger space yet remains weaker even after 20 generations. By seeding layer-wise NSGA-II with type-wise candidates, the guided search reaches a stronger Pareto frontier within 5 generations: for the near-clean 88\% target, it reduces extra cost from 1.73 to 1.56, a 10.3\% reduction, and increases feasible choices from 10 to 18, providing richer deployment options. 
% These results show that layer-type sensitivity is an effective prior for navigating the heterogeneous encoding space and improving the robustness-cost trade-off.

\begin{figure}[!t]
\centering
\includegraphics[width=\columnwidth]{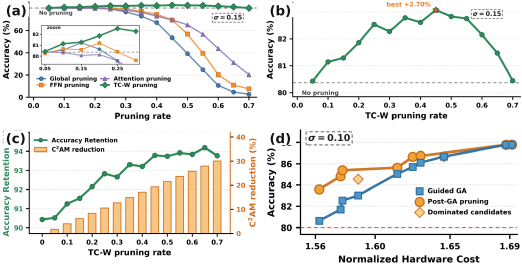}
\caption{Post-GA pruning as robustness refinement.}
\label{fig:post_ga_pruning}
\end{figure}

% \textbf{Post-GA pruning further improves robustness by removing harmful small-magnitude TC-encoded weights.}
Fig.~\ref{fig:post_ga_pruning} evaluates post-GA pruning as a robustness refinement for selected deployment candidates. Pruning TC-W layers is the most effective scope because small-magnitude negative weights in these layers can still introduce sign-extension ones and high-significance conducting events. The pruning-ratio sweep improves robustness up to an intermediate ratio, with the best setting giving a 2.70\% accuracy gain over no pruning at \(\sigma=0.15\); beyond this point, excessive pruning damages clean model capacity. Fig.~\ref{fig:post_ga_pruning}(c)--(d) links the gain to C\(^2\)AM reduction and shows that pruning lifts the Pareto frontier at unchanged normalized cost.

\begin{figure}[!t]
\centering
\includegraphics[width=\columnwidth]{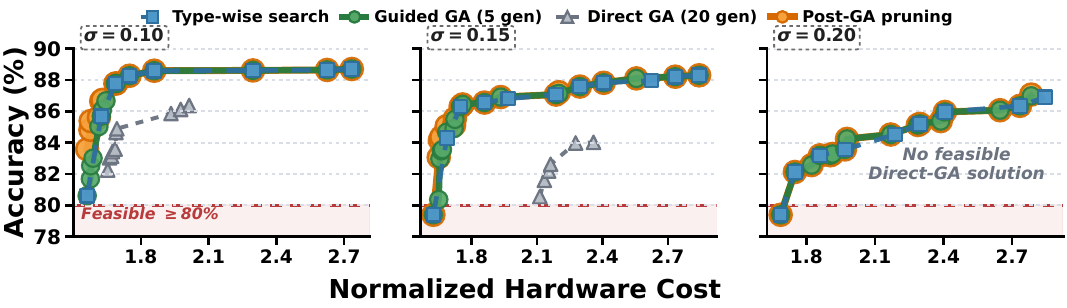}
\caption{Accuracy-cost Pareto trends across FeFET variation strengths.}
\label{fig:pareto_across_sigma}
\end{figure}

% \textbf{Guided heterogeneous deployment remains effective across variation strengths.}
Fig.~\ref{fig:pareto_across_sigma} compares Pareto frontiers across three FeFET variation levels. Type-wise search quickly provides feasible starting points but is constrained by shared layer-type assignments, while direct layer-wise NSGA-II converges less effectively and fails to meet the 80\% feasibility threshold at \(\sigma=0.20\). The sensitivity-guided NSGA-II consistently achieves stronger frontiers by combining type-level priors with layer-wise flexibility, and post-GA pruning further improves accuracy at unchanged cost. These results show that adaptive encoding and pruning remain complementary across variation strengths.

\begin{figure}[!t]
\centering
\includegraphics[width=\columnwidth]{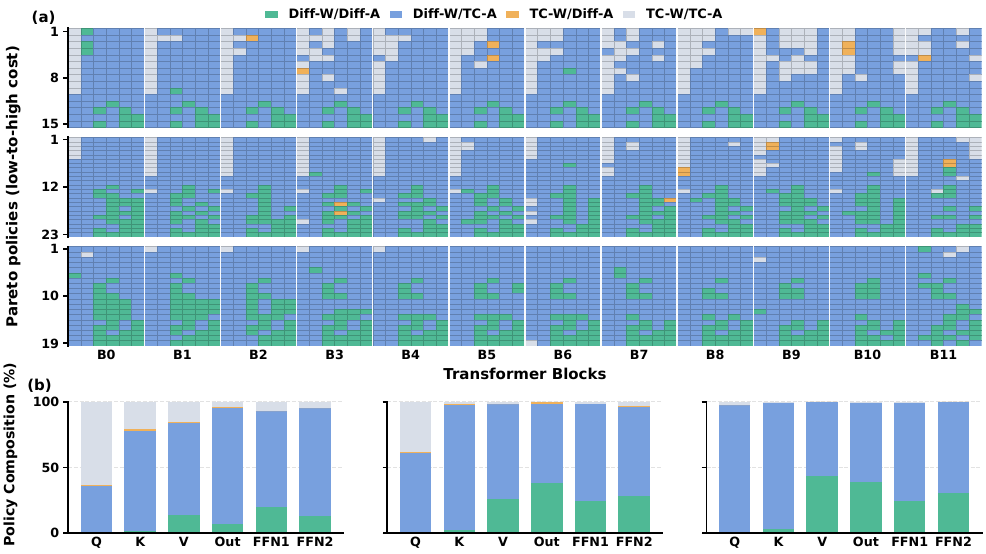}
\caption{Selected adaptive encoding policies and layer-type composition.}
\label{fig:selected_policy_composition}
\end{figure}

% \textbf{CAFE learns structured heterogeneous policies instead of applying differential encoding uniformly.}
Fig.~\ref{fig:selected_policy_composition} shows structured adaptive policies rather than uniform full-differential protection. As the variation strength increases, the selected policies use stronger differential encoding, but value and output projections are protected more often than the more sensitive FFN layers because their lower hardware cost improves the accuracy-cost trade-off. This confirms that CAFE allocates differential protection according to both robustness sensitivity and hardware cost.
% These patterns demonstrate that CAFE allocates differential encoding according to both robustness sensitivity and hardware cost, yielding structured deployments that are more efficient than uniform full-differential protection.

\section{Conclusion}

This paper presented CAFE, a conducting-cell-aware adaptive differential encoding framework for robust FeFET-based CIM inference. By exposing how signed encodings amplify conducting-cell variation in bit-sliced FeFET CIM, CAFE introduces C\(^2\)AM to quantify bit-significance-weighted conducting activity and guide differential encoding, layer-wise adaptive deployment, and post-GA pruning. CAFE retains model performance while limiting hardware overhead: across representative workloads, it reduces C\(^2\)AM by up to 28.6\(\times\), improves accuracy under variation by more than 87 percentage points relative to two's-complement CIM, and approaches full-differential robustness at much lower normalized cost.

\section*{Acknowledgment}

This work was supported by National Key R\&D Program of China (2022YFB4400300), NSFC (62534002, 62274003, 92564302) and 111 Project (B18001), in part by the STIC under Grant QYJS-2022-1501-B and the National Micro/Nano Fabrication Laboratory of Peking University.

\bibliographystyle{IEEEtran}
\bibliography{refs}

\end{document}